**Advanced sine wave modulation of continuous wave laser system for atmospheric $CO_2$ differential absorption measurements**

Joel F. Campbell, Bing Lin, Amin R. Nehrir
NASA Langley Research Center
Hampton, VA 23681
joel.f.campbell@nasa.gov

## Abstract

In this theoretical study, modulation techniques are developed to support the Active Sensing of $CO_2$ Emissions over Nights, Days, and Seasons (ASCENDS) mission. A CW lidar system using sine waves modulated by ML pseudo random noise codes is described for making simultaneous online/offline differential absorption measurements. Amplitude and Phase Shift Keying (PSK) modulated IM carriers, in addition to a hybrid pulse technique are investigated that exhibit optimal autocorrelation properties. A method is presented to bandwidth limit the ML sequence based on a filter implemented in terms of Jacobi theta functions that does not significantly degrade the resolution or introduce side lobes as a means of reducing aliasing and IM carrier bandwidth.

## Introduction

NASA Langley Research Center in collaboration with ITT Exelis has been evaluating Continuous Wave (CW) laser absorption spectrometer (LAS) systems as a means of retrieving atmospheric $CO_2$ column amounts from space to support the ASCENDS mission. The U.S. National Research Council's report entitled Earth Science and Applications from Space: National Imperatives for the Next Decade and Beyond identified the need for an active lidar space mission. This report targeted a need for measurements of column $CO_2$ to gain a better understanding of $CO_2$ sources and sinks through the ASCENDS mission[1]. The primary objective of the ASCENDS mission is to make $CO_2$ column mixing ratio measurements during day and night over all latitudes and all seasons and in the presence of thin or scattered clouds. These measurements would be used to significantly reduce the uncertainties in global estimates of $CO_2$ sources and sinks, to provide an increased understanding of the connection between climate and atmospheric $CO_2$, to improve climate models, and to close the carbon budget for improved carbon and chemical weather forecasting and as well as implementing policy decisions [1].

The critical component in the ASCENDS mission is the Laser Absorption Spectrometer (LAS) for $CO_2$ column measurements with weighting towards the mid to lower troposphere using the Integrated Path Differential Absorption (IPDA) technique from the surface and cloud backscattered returns.

IPDA lidar systems measure the difference of the total gas absorption along the path length between two or more laser wavelengths in the path of the laser to a scattering target. For the earth's climate system, there are large changes in the atmospheric variables such as cloud and aerosol distributions, and the profiles of water vapor and other absorbing gases from regional to global scales. Also, the surface reflectance, varies with many factors and variables such as surface type, roughness, slope, vegetation, and soil moisture. The variations in these atmospheric and surface variables can generate large changes in received lidar powers from the surface or other targets, which can introduce large biases and artifacts in determination of $CO_2$ column amounts when only a single laser with a single wavelength is utilized. The advantage of IPDA LAS system is that the strong effects of the variations in atmospheric states and surface reflection on lidar return powers would be minimized from the ratio of the simultaneously received lidar powers from the two closely spaced online and offline IPDA wavelengths. All the influences on the lidar return signal from these environmental variables are the same for the two or more closely spaced wavelengths. The ratio tends to minimize the environmental effects except the differential absorption optical depth at the desired LAS wavelengths. Thus, with all kinds of the considerations including experimental factors, atmospheric temperature and pressure profiles, and range estimations, high accuracy $CO_2$ column measurements can be achieved from the $CO_2$ differential optical depth.

For the purposes of this paper we consider the two-wavelength approach, which uses one laser with the wavelength tuned to the $CO_2$ gas absorption line (on-line), and the other wavelength tuned away from the absorption feature but still within 100 pm from that same absorption line (off-line). The differential absorption optical depth for the $CO_2$ column may be derived from the ratio of the received optical power from each wavelength. There are several different approaches to making this measurement.

$CO_2$ IPDA LAS measurement approaches include pulsed [2-3] and CW systems. Direct detection pulsed systems use a technique where measurements for each of the IPDA wavelengths are sampled successively via sequential laser shots sampled through different atmospheric columns and surface spots at the surface. The CW approach described here transmits the online and offline lidar signals to the surface simultaneously from a single laser transmitter, thereby allowing for simultaneous sampling of the atmospheric column background noise, and surface reflectivity within the laser spot, and thus, reducing the potential errors. In addition, a major advantage in this approach is the significant reduction of low frequency variations in background radiation and colored noise, which are filtered out in a natural way with IM carrier-based matched filter approaches by effectively limiting the detection frequency and bandwidth. This is done through the demodulation process shown in Fig. 2, which limits the frequency range to a band of frequencies about the carrier frequency, thus filtering out unwanted low frequency noise. Within the CW approach, IM techniques are used to separate the simultaneous return signals from the different IPDA wavelengths, provide ranging capability, and discriminate surface returns from other atmospheric returns such as aerosols and clouds. The IM technique is also useful for extracting the laser signals from solar background noise, which occurs as a byproduct of the matched filter technique used to retrieve the science signal of interest. Potential IM-CW techniques include pseudo-noise (PN) code CW [5-6], linear swept sine wave CW [7-12], unswept sine wave CW [4, 13-16], non-linear swept sine wave CW [17], etc. The baseline technique used by NASA Langley Research Center in partnership with ITT Exelis for the ASCENDS mission is linear swept frequency CW [12]. The utility of this technique for both range and $CO_2$ column absorption measurements has been demonstrated over the past several years through ground-based and airborne demonstrations [12].

A drawback of various modulation techniques, such as linear swept frequency, is the potential for generation of side lobes in the autocorrelation function, which could introduce small bias errors when intermediate scatterers such as thin clouds are close to measurement target. This is the main advantage of digital techniques such as PN codes and non-linear frequency modulation. Previously, PN codes have been described using a pure maximum length (ML) sequence with time shifting to separate channels [5]. This technique may have the advantage of eliminating the side lobe issue for certain hardware configurations in LAS systems, while allowing for multiple channels that share the same bandwidth. Additionally, the time shifting approach allows for simple and straight forward data processing of multi-channel laser returns since only a single matched filter correlation needs to be performed for all channels instead of a separate matched filter correlation for each channel. One potential disadvantage with this technique is that not all hardware configurations have a suitable low frequency response for transmitting and/or receiving a pure ML sequence without distortion. Since a ML sequence has a main power band at zero frequency, using such a technique in band limited hardware (high pass or band pass filtered in particular) will filter out the main power band and corrupt the autocorrelation properties, which can result in non-orthogonality between channels in a time shifted multi-channel configuration. This problem could be addressed by imparting a PN code amplitude modulation on a narrow band carrier at a predetermined frequency, performing amplitude demodulation, and conducting matched filter correlation with the reference PN code to obtain the range capability. A low cost demonstration of this was implemented using Hilbert Transforms forms for amplitude demodulation. This demonstrated much better autocorrelation properties than linear swept frequency but still exhibited minor artifacts, and the demodulated PN code exhibited minor distortions and ringing [6]. This was a student demonstration implemented in audio and was instrumental in demonstrating remote sensing concepts using PN codes on a carrier, and was also a very nice demonstration outside of the classroom as well. In this paper we eliminate those minor issues by implementing an improved demodulator. We also introduce improved modulation techniques beyond amplitude modulation, which improves Signal to Noise Ration (SNR). To further simplify our swept frequency IM-CW LAS system for future a ASCENDS mission, we concurrently investigate advanced IM techniques, and develop modulation/demodulation techniques that are simple and have autocorrelation properties that rival those of the pure ML sequence. This includes a quadrature correlator demodulator, PSK modulation techniques, and hybrid pulse modulation. This results in near perfect autocorrelation properties and an increase in SNR by up to a factor of 2.

Different modulation techniques are discussed in following sections appropriate for the hardware shown in Fig. 2. Key results are summarized in the conclusion. Details of the actual instrument, atmospheric models, etc. have been discussed elsewhere [12, 18] and will not be repeated here.

**Measurement technique**

The concept of our IM-CW system for $CO_2$ IPDA measurements is illustrated in Figure 1. For the system described in this figure, amplitude modulated online and offline seed lasers with good spectral properties are optically combined using fiber optics and used to simultaneously seed a single Erbium Doped Fiber Amplifier (EDFA) to increase the transmitted power. A small fraction of the transmitted beam is picked off via an optical tap inside of the EDFA and sent to a reference detector for energy normalization. The backscattered science signal from the online and offline wavelengths from the surface as well as aerosols and clouds are simultaneously collected with a telescope, optically filtered with a narrow band optical filter, and detected by a single detector. Both the science and reference signals are amplified, electronically filtered and then digitized for retrievals of column $CO_2$. Post processing of the digitized science and reference data allow for discrimination between ground and intermediate scatterers (from clouds, etc.) using the matched filter technique, and also to obtain differential absorption power ratios for inference of $CO_2$ column amounts as well as range estimates to the scattering targets.

The on-line and off-line laser modulation signals using the time shifted approach for this study can be expressed as

$$\begin{aligned} \Lambda_{off} &= 1 + m\xi(t), \\ \Lambda_{on} &= 1 + m\xi(t - \Delta t), \end{aligned} \tag{1}$$

where m is the modulation index with a value between 0 and 1, and $\xi(t)$ is the repeating modulation waveform where $-1 \le \xi(t) \le 1$, $\langle \xi(t) \rangle = 0$, and $\Delta t$ is the time shifting of online modulation waveform related to that of offline. In general, we choose $\Delta t > 2r_{\max} / c$ to avoid interference between the time shifted channel and intermediate scatterers, where $r_{\max}$ is the maximum distance among potential scatterers including the Earth's surface and is also called unambiguous range, while c is the speed of light.

Given the modulation waveforms presented in Eq.1, and using the relation $P^T_{on/off}(t) = \overline{P^T_{on/off}} \Lambda_{on/off}(t)$, the instantaneous received lidar off-line and on-line received optical powers ($P^R_{off}(t)$ and $P^R_{on}(t)$, respectively) from a single scattering target (i.e. the surface) at range r (which may also vary with time depending on the experiment) are given by

$$P_{off}^{R}(t)=\frac{K}{r^{2}}\,\overline{P_{off}^{T}}\exp\left(-2\varepsilon\int_{0}^{r}\beta(r')dr'\right)\exp(-2\tau)\left(1+m\xi(t-2r/c)\right)$$
$$P_{on}^{R}(t)=\frac{K}{r^{2}}\,\overline{P_{on}^{T}}\exp\left(-2\varepsilon\int_{0}^{r}\beta(r')dr'\right)\exp(-2\tau)\exp(-2\tau')\left(1+m\xi(t-2r/c-\Delta t)\right),\quad (2)$$

where $\tau$ is the total one-way column optical depth for the off-line and on-line measurement that is determined by the gas absorption from molecules other than $CO_2$, and $\tau'$ is the one-way column optical depth resulting from $CO_2$ absorption only, $\beta$ is the backscatter coefficient ($km^{-1}sr^{-1}$), $\varepsilon$ is the extinction to backscatter ratio ($sr^{-1}$), K is a constant, where $\overline{P_{off}^{T}}$ and $\overline{P_{on}^{T}}$ are the average transmitted on-line and off-line powers averaged over one period of the modulation waveform, respectively.

When multiple targets including the surface exist in the lidar path length, each target would generate similar received lidar signals like those in Eq. 2. The signals from these multiple targets are combined at the detector and converted to an electronic signal *S*(*t*). For an AC coupled receiver subsystem, we have,

$$S(t)=\sum_{k}\left[C_{1k}m\xi(t-2r_{k}/c-\Delta t)+C_{2k}m\xi(t-2r_{k}/c)\right],\quad (3)$$

where we sum over all k scatterers and where,

$$C_{1k}=\frac{K_{k}'}{r^{2}}\,\overline{P_{on}^{T}}\exp\left(-2\tau\int_{0}^{r_k}\beta(r')dr'\right)\exp(-2\tau_{k})\exp(-2\tau_{k}')$$
$$C_{2k}=\frac{K_{k}'}{r^{2}}\,\overline{P_{off}^{T}}\exp\left(-2\varepsilon\int_{0}^{r_k}\beta(r')dr'\right)\exp(-2\tau_{k}),\quad (4)$$

where K' is a constant. These $C_{1k}$ and $C_{2k}$ returns can be uniquely discriminated from other returns from different scatterers using a matched matched filter, which involves correlating the received waveform with a reference signal. These are also proportional to the average received optical power ($\overline{P_{off,\,k}^{R}}$ and $\overline{P_{on,\,k}^{R}}$) from the kth scatterer. For a ground target solving for $\tau_{g}'$ gives

$$\tau_{g}'=\frac{1}{2}\ln\left(\frac{C_{2g}\overline{P_{on}^{T}}}{C_{1g}\overline{P_{off}^{T}}}\right)=\frac{1}{2}\ln\left(\frac{\overline{P_{off,\,g}^{R}}\,\overline{P_{on}^{T}}}{\overline{P_{on,\,g}^{R}}\,\overline{P_{off}^{T}}}\right),\quad (5)$$

where $C_{1g}$ and $C_{2g}$ are the ground returns. Once $C_{1g}$ and $C_{2g}$ are determined, the column optical depth for $CO_2$ can be found using equation 5. In general, this is done by cross correlating the reference waveform with the return signal *S*(*t*), i.e., by the matched filter technique as mentioned before. This results in multiple peaks for the multiple scatterers including the surface. The ground return can be identified through range gating (or maximum range). By choosing the reference modulation waveforms with perfect autocorrelation properties (ML sequence for instance), channel separation is achieved through the time shifting approach since each channel correlates at a different apparent range without interference between channels or scatterers. That is, each scatterer results in two peaks in its matched filter output: one for the offline returns, and the other for online returns with the $\Delta t$ shifted time. The details of this approach are discussed elsewhere [5].

**Band pass filtered modulation waveform generation and demodulation**

We now present the improved modulation approach for $CO_2$ IPDA measurements. This new method modulates an IM sine wave carrier by an ML-sequence as in Equation 6. To effectively represent each bit of ML-sequence in the transmitted waveforms, and allow flexibility in the unambiguous range for a particular code length and sample rate, we oversample the ML-sequence by an integer *M* such that each code bit of the ML sequence is represented by *M* points. We then use that bit to modulate a sine wave that has exactly *P* cycles per code bit. In order to satisfy the Nyquist rule we must have $M > 2P$ . Let $z_n \equiv z(n)$ be the original ML-sequence and $Z_n \equiv Z(n)$ be the oversampled ML-sequence such that $Z(n) = z(\mathrm{int}((n-1)/M)+1)$, where int(x) is a function that represents the integer part of x. These definitions are used in the following sections wherever PN code modulation is discussed. The following two subsections describe amplitude and PSK IM carrier modulation and demodulation techniques. Amplitude modulation is achieved by modulating the IM carrier with an ML sequence. With PSK modulation we use the ML sequence to shift the phase of the IM carrier by 180 degrees, which is equivalent to changing the sign of the IM carrier. The PSK modulation technique makes more efficient use of available optical amplifier power and results in better signal-to-noise ratio (SNR).

<u>IM carrier amplitude modulation</u>

In the case of amplitude modulation, the modulation waveform takes on one of two forms.

$$\begin{aligned} \xi_a(n) &= Z(n)\cos(2\pi(n-1/2)P/M), \\ \xi_b(n) &= Z(n)\sin(2\pi(n-1/2)P/M). \end{aligned} \tag{6}$$

The ½ factor is introduced to make the autocorrelation function symmetric.

Demodulation is accomplished as in Figure 2 using the reference kernel of $2Z(n)$-1. This particular reference kernel is necessary to get a resulting autocorrelation function with off main lobe values of zero with the amplitude modulation described by Equation 6. That is to say, with this reference kernel the autocorrelation function is zero except at the main lobe. This is a quadrature demodulator [19] with the low pass filter replaced by a matched filter correlation with the oversampled ML-sequence. The correlations can be calculated using a DFT, which are most efficiently computed with a Fast Fourier Transform (FFT) and take the form [5,6]

$$
\begin{aligned}
R(ref, data) &= \frac{1}{N}\sum_{m=0}^{N-1} ref^*(m)\ data(m+n) \\
&= DFT^{-1}\left(DFT^*(ref)\,DFT(data)\right),
\end{aligned} \tag{7}
$$

where "*" denotes the complex conjugate, N is the length of oversampled ML-sequence, ref is defined as the reference signal, and data is defined as the received data.

As an example of a modulation technique that is suitable for a band pass filtered LAS system presented in figure 1, we take the special case where P=1 (one cycle per code bit) and M=8 (8 samples per code bit). We generate an 8$^{th}$ order (or 255 length) ML-sequence using the recurrence relation $z_{n+8} = z_n \oplus z_{n+2} \oplus z_{n+5} \oplus z_{n+6}$ with the seed {1, 0, 1, 0, 1, 1, 1, 1}, where $\oplus$ is the "exclusive or". We then oversample this by a factor of 8. Figure 3 shows a comparison between the cosine wave vs sine wave modulation as in Equation 6. Note that each waveform in this and all subsequent plots are plotted with 8 samples per code bit with linear interpolation between points.

The autocorrelation function may be computed according to the block diagram in Figure 2. The autocorrelation function for both the cosine and sine wave cases are shown in Figure 4, using a 35 sample delay. For larger values of P the autocorrelation for both the sine and cosine modulation cases approaches that of the pure PN code autocorrelation function.

It can be shown that the half height width for both the cosine and sine wave cases is identical to the pure ML-sequence case, though the sine wave case may be better suited for absorption measurements because of the more rounded (flatter) peak. The maximum value for an autocorrelation function with a very sharp peak will show sensitivities to under sampling and therefore may result in a bias. The reason for this is because the return signal will be delayed in such a way that the exact peak location of the autocorrelation function may not always fall on a sample point. As a result there will be a range dependent jitter in the height measurement. If the peak is sharp the jitter will be worse than if it is relatively flat. Another reason is that typically one uses some sort of interpolation to get a better estimate of the peak height and this is much easier to do if the peak is a continuous function. The sine wave autocorrelation function is less sharp than the pure ML-sequence. In both cases and within the accuracy of numerical calculations, the off main lobe values are essentially zero with the chosen parameters and normalization. The resolution of discriminating different targets is close to the pure ML-sequence case of $c/2B_r$ [5,6], where $B_r$ is the code bitrate (number of code bits per second), which is related to the effective modulation bandwidth. It should be noted the autocorrelation function in the case of more than 2 cycles per code bit results in an autocorrelation function that looks very similar to the pure ML sequence case [5,6].

PSK IM carrier modulation

Our particular PSK modulation takes one of two forms.

$$\xi'_a(n) = \left(2Z(n)-1\right)\cos\left(2\pi\left(n-1/2\right)P/M\right),$$
$$\xi'_b(n) = \left(2Z(n)-1\right)\sin\left(2\pi\left(n-1/2\right)P/M\right). \tag{8}$$

As in the previous case, the ½ factor is introduced to make the autocorrelation function symmetric. This is very similar to the amplitude modulation case except that $Z(n)$ has been replaced by $2Z(n)$-1. Figure 5 shows a comparison between the cosine wave vs sine wave modulation as in Equation 8.

Demodulation is accomplished as in Figure 2, which results in the best autocorrelation properties. Another PSK demodulation technique sometimes used in radar is to correlate the outgoing PSK modulated sine wave signal with the reference PSK modulated sine wave, but that results in less than ideal autocorrelation properties that contain side lobes [20]. In this case, we use the reference kernel $Z(n)$ with aforementioned demodulator in Figure 2. This reference kernel is necessary to get a resulting autocorrelation function with off mainlobe values of zero in the case of PSK. The autocorrelation function is identical to that shown in Figure 4 in the case where P=1 (one cycle per code bit) and M=8 (8 samples per code bit). In spite of identical autocorrelation functions, this case has a major advantage. Since we are using $Z(n)$ as a kernel, this method has better noise immunity. The reason is because $Z(n)$ is 0 for half of the length, which means we correlate over half the noise compared to the amplitude modulated case. As a result the SNR should be better a factor of $\sqrt{2}$ since the amplitude for each is the same. However, if we were to use $2Z(n)$-1 for the kernel in the PSK modulation case the SNR would be even better. The noise increases by a factor of $\sqrt{2}$ but the amplitude increases by a factor of 2. The net result is that the SNR is a factor of 2 better than the amplitude modulation case. These results are explained in detail in Appendix B. The downside is there is a slight constant bias in the background that is inversely proportional to the bit length of the code [5], which makes it undesirable for the amplitude measurements necessary for $CO_2$ retrievals. This should present little or no problem for range measurements, however. As a result the $Z(n)$ kernel version used here is the likely candidate for amplitude measurements because of the superior autocorrelation properties.

**Low pass filtered modulation waveform generation and demodulation**

<u>Hybrid pulse/CW modulation</u>

In lidars designed to detect pulses such as those with photon counting receivers or other types of DC coupled receiver subsystems, other types of modulation are possible. For an alternative implementation, we consider the case where we have DC coupled electronics, which for the purpose of this paper is achieved by using the low pass filtered configuration in Figure 1, combined with an EDFA in the transmitter. An EDFA amplifier utilizes an internal pump laser with a wavelength typically at 980 nm that continually pumps the erbium atoms in the erbium doped fiber to an excited state. An external signal laser (which in our case has a wavelength of approximately 1570 nm) is used pass photons through the erbium doped fiber. Excited erbium atoms then give up some of their energy at the same wavelength and phase as the signal laser, which has the effect of amplifying the laser signal. There are two types of modulation possible with an EDFA amplifier. The first is pump modulation. Modulating the intensity of the pump laser produces a change in gain of the signal, which modulates the output of the EDFA. With pump modulation there is a low pass filter response [21]. This would be an ideal configuration but it requires an optical amplifier for each channel. However, with signal modulation as has been used in the past [12], which allows one to use the same amplifier for all channels, this system behaves as a lead compensator circuit [21], which is similar to a high pass filter. This is accomplished by modulating the input laser signal directly. In our case we modulate this signal by varying the intensity. The step response of such an amplifier configured for signal modulation, which is shown in Figure 6, does not fully follow the input step function. This is due to an excited state population depletion, which then settles to a steady state.

Even if the transmitter and receiver electronics are DC coupled, the nature of the optical amplifier used here conspires to degrade the autocorrelation properties due to the distortion of the PN code. By modulating the PN code by an offset carrier, one can potentially avoid those sorts of distortions because the optical amplifier responds better to quick changes in signal input than it does to maintaining a constant level as is required by a pure PN code. One solution is to build this a pulse train into the modulation signal. One could do this with a square wave if one has enough bandwidth. A particularly simple method would be to modify the Equation 6 to

$$\begin{aligned}\xi(n) &= \left(1-\cos\left(2\pi(n-1/2)P/M\right)\right)Z(n)-1 \\ &= 2\sin^2\left(\pi(n-1/2)P/M\right)Z(n)-1\end{aligned} \qquad (9)$$

An example of this modification is shown in Figure 7 with the first 100 points of the modulation waveform. For this example, P=1 and M=8 are selected. This modification has one main advantage over the previous technique: for the same amplitude it uses only half the average optical power. The reason is because in order to transmit the previously described modulation it must be offset. This means that a logical zero in that case is half the peak laser power whereas in this case a logical zero is zero power.

Demodulation of this is particularly simple. Instead of demodulating the result using the block diagram in Figure 2, one simply has to correlate the return signal with $2Z(n)$-1. The resulting autocorrelation function is shown in Figure 8.

As in the previous case, the half height width is identical to the pure ML-sequence modulation. However, due to the more rounded peak of the autocorrelation function, the height measurement, which is desirable for accurate $CO_2$ absorption measurements, should be less sensitive to sampling than with the sharp peak of the pure ML-sequence case or the ML-sequence modulated sine wave case in the previous section. Yet the autocorrelation properties are near perfect as those in previous cases with the chosen parameters and normalization.

**Bandwidth limited codes**

In most LAS implementations it is desirable to limit the bandwidth of the PN code to reduce aliasing and thereby minimize sampling error in the measurement of the amplitude height of the autocorrelation function. This is particularly important in a space application because there is an upper limit to how fast one can sample. For instance as of the time of this writing, the fastest one can digitize with space qualified parts, at 16 bit resolution is 10 MHz, that essentially sets the upper limit for the spatial distance between points at about 15 meters. Since we need 3 meters resolution accuracy to get 1-ppm accuracy that puts severe limits on what we can do without interpolation and interpolation doesn't work well with an aliased signal. Since we can't sample at any rate we want to avoid aliasing, the only option left is to band limit the signal somehow. This helps with the absorption measurements and makes the peak height less sensitive to small changes in range. The question is how to accomplish this requirement? Many filters can introduce sidelobes to the autocorrelation function, such as a square window filter or some other high order filter. This also potentially includes a non-optimal band pass filter, which may already be present so filtering the PN code in the correct way before hand in combination with putting the PN code on a carrier can prevent the problem. Gaussian filters are sometimes used in digital communications to bandwidth limit digital data, minimize channel bleed-over, and carry the maximum amount of information with the least bandwidth [22]. A convenient implementation is to use a periodic version of a Gaussian filter so that we can terminate the filter at the end of a cycle. Since a Discrete Fourier Transform (DFT) contains the frequency content in the first half of the array, and a mirror image in the second half, and the PN code itself is transmitted as a repeating waveform, a cyclic filter convolution is appropriate.

We may construct a periodic Gaussian in the following way with

$$G(x) = \sum_{n=-\infty}^{\infty} \exp\left((x-nL)^2 / 2\sigma^2\right). \tag{10}$$

Here L is the period, $\sigma$ is the standard deviation, x is an as yet undefined variable, and L is the period . This sum can be done analytically (see appendix) as

$$\sum_{n=-\infty}^{\infty} \exp\left((x-nL)^2 / 2\sigma^2\right) = \sqrt{2\pi}\left|\frac{\sigma}{L}\right|\left[1+2\sum_{n=0}^{\infty} q^{n^2} \cos(2\pi nx / L)\right] \equiv$$
$$\sqrt{2\pi}\left|\frac{\sigma}{L}\right| \vartheta_3(\pi x / L, q) \quad , \tag{11}$$

where $\vartheta_3$ is the Jacobi theta function of the third kind and the nome q is given by

$$q = \exp\left(-2\pi^2\sigma^2 / L^2\right). \quad (12)$$

Theta functions are widely used in the theory of elliptic functions [23] and other applications [24-25]. In order to scale this so that it is 1 at the maxima we have

$$W(x, L, q) = \vartheta_3(\pi x / L, q) / \vartheta_3(0, q) \quad . \quad (13)$$

To apply this as a filter in the frequency domain we simply multiply the DFT by the array $W(n-1, N, q)$ and take the inverse DFT, where N is the length of the array and $1 \le n \le N$. Here we have taken $x$=$n$-1 with period $N$ since the analytic continuation of a DFT has period $N$ if $n$ is the array index.

To demonstrate the usefulness of this filter we first show how to limit the bandwidth of an ML-sequence and how the autocorrelation properties and resolution are preserved. We then use that bandwidth limited sequence in the two modulation examples discussed in the previous sections to show both the effect of bandwidth limiting and preservation of the near perfect autocorrelation properties for each case with only a minor degradation in resolution. Because of these characteristics, the discussed modulation codes could be easily implemented in current $CO_2$ LAS systems and the concept of future space instruments.

Bandwidth limited ML-sequence

We start with the consideration of the simple case of a pure ML sequence with length 255 and oversampled by a factor of 8 as in the previous section. We filter this with the theta function filter with $q$=0.9 and N=255×8=2040. A plot of the unfiltered vs. filtered PN code frequency content is shown in Figure 9.

If the original PN code modulation were applied to a LAS system with a non-optimal filter, aside from the spread in the autocorrelation function, the spectrum content could be distorted in a way that could introduce possible artifacts such as side lobes. By bandwidth limiting the codes, it is possible for the modulation bandwidth to fit well within the bandwidth of the band pass or low pass receiver filter such that the receiver filter impact is minimal to virtually eliminate the side lobe issue, which has been confirmed by simulations. These concepts point out practical ways to implement our PN codes to current and future $CO_2$ LAS instrumentation.

The waveform shapes of these unfiltered vs. filtered PN codes are shown in Figure 10. The waveform of the filtered code is obtained from the circular convolution between the cyclic theta function filter and the cyclic PN code.

The results of the correlation between 2$Z$($n$)-1 and both the filtered and unfiltered PN codes are shown in Figure 11. In spite of the strong filtering, the main lobe width of the autocorrelation function is very similar, and the peak is more rounded, which is a desirable trait for accurate measurements of $CO_2$ column amounts. Also, if the waveform is sampled at higher than the Nyquist rate, the peak height of the autocorrelation function can be determined more accurately through interpolation because the continuous function it is based on can always be reconstructed exactly. Shannon's version of the Nyquist theorem states exactly that [26]. Of course, noise and other errors would tend to make this process less exact. But it does show that bandwidth limiting the waveform to below the Nyquist sample rate is desirable if one is interested in absorption measurements. Because of the required IM carriers for PN modulation in the instrumentation as discussed previously, we further investigate the characteristics of sine wave and hybrid modulations for the filtered PN codes to ensure the usefulness of current discussed techniques for LAS applications.

<u>Amplitude modulated IM carrier using theta function filtered PN code</u>

In this case we use the filtered PN code in previous subsection to modulate a sine wave IM carrier as in $\xi_b$ of Equation 6 with P=2 (2 cycles per code bit) and M=8 (8 samples per code bit). A plot of the resulting modulation is shown in Figure 12.

The frequency spectra of the modulated sine wave are plotted in Figure 13. This shows that the resulting frequency content of the filtered modulation is bounded within the Nyquist range with no aliasing.

The autocorrelation functions of these filtered vs unfiltered PN modulations are illustrated in Figure 14. This shows the modulation produces more robust matched filter output for peak determination by producing a more rounded peak, which is desirable for amplitude measurements in differential absorption estimation because it produces less sampling error. The resolution degradation is very minor.

<u>PSK modulated IM carrier using theta function filtered PN code</u>

In this case we use the filtered PN code described earlier to modulate a sine wave IM carrier as in $\xi'_b$ of Equation 8 with P=2 (2 cycles per code bit) and M=8 (8 samples per code bit). A plot of the resulting modulation is shown in Figure 15.

Figure 16 is a plot of the frequency spectra of the modulated sine wave. This shows the resulting frequency content of the filtered modulation is also bounded within the Nyquist range with no aliasing as in the amplitude modulation case.

The plot of the autocorrelation function of the filtered vs unfiltered PN PSK modulation is identical to Figure 14 of the previous section.

<u>Hybrid pulse modulation with theta function filtered ML sequence</u>

We take the case using Equation 9 with P=1 (one cycle per code bit) and M=8 (8 samples per code bit), and use the same filtered PN code as in the previous two examples with q=0.9. Figure 17 shows the resulting modulation.

The Resulting frequency spectra are shown in Figure 18. Here we see the original modulation with the chosen parameters already does a good job at limiting the high frequencies as one might deduce from Figure 6. This is because there are no sharp transitions as there are with the pure ML-sequence. The filtered version limits the spectra even more.

Figure 19 is a plot of the autocorrelation function of the filtered vs unfiltered PN code modulation. Both cases show very good autocorrelation properties.

**Discussion and summary**

Table 1 shows a summary of the various modulation techniques covered in this study. We have presented several different PN modulation techniques that can be used with different types of hardware: the baseline system in Figure 1 with a signal modulated EDFA, the baseline with the lowpass filter configuration, or the baseline system with the lowpass configuration and using separate EDFAs with pump modulation. In the case of DC coupled electronics combined with a laser amplifier with pump modulation, a pure or filtered ML sequence not on an IM carrier is ideal. In the case of the baseline, although the amplitude modulated sine wave version works well, the PSK modulation case discussed here has superior signal to noise. If the electronics is DC coupled but the laser has a transient response such as the case with a signal modulated EDFA, the hybrid pulse version is best. In addition, a new type of cyclic filter based on the Jacobi theta function was introduced, which is useful for bandwidth limiting digital PN codes without seriously degrading the resolution and resulting in autocorrelation properties completely without side lobes or other artifacts. The main benefit is to reduce aliasing, which reduces sampling error when measuring the amplitude of the autocorrelation function. It can also minimize the impact of a non-optimal filter by minimizing side lobes caused by that filter as has been confirmed by numerical simulations with filters used in previous experiments. By using a combination of IM carrier modulation and theta function filtering, one can design a modulation scheme that is flexible enough to work with most hardware while maintaining the integrity of the autocorrelation properties. Internal studies have shown the previous linear swept frequency technique, which exhibits a sidelobe level of up to 13 dB, used in our previous test range and flight campaigns has shown a degradation in accuracy due to sidelobe interference from thin clouds. Although one could potentially develop numerical techniques to deal with this issue, it is much simpler to eliminate the problem by using a modulation and demodulation that doesn't have sidelobes. This helps meet ASCENDS program requirements by reducing computational burden. Another benefit that has been discussed in a previous publication [5] is the use of the time-shifted approach to separate channels. That approach works with every modulation technique presented here and has a major advantage in that the demodulation process can be done with a single demodulation step for all channels. This further simplifies the computational burden when compared to the linear swept frequency technique, which requires a separate demodulation for each channel. Numerical simulations based on a previous model for the system used up to this point [18] show that these techniques could potentially benefit the ASCENDS program with little or no change to the existing baseline hardware and will be explored in a future experimental study.

**Appendix A: SNR for AM, PSK, PN, and Hybrid Pulse Modulation**

Lets suppose we have a sine wave signal amplitude modulated by a PN code, Z with amplitude a in the presence of white noise (for example) $\eta$ with variance $\sigma^2$ and 0 mean. For simplicity we look at the pure electronic signal as pure signal plus white noise. Note that in reality the noise variance will also be a function of the average received optical power if we have detector shot noise, which is normally modeled as white noise. If we assume equal average power this simple white noise with variance $\sigma^2$ assumption is valid. This is sufficient for determining whether amplitude or PSK modulation has better SNR, and what it is under those assumptions.

Noise through the quadrature correlator

2Z-1 kernel case:

We first look at the noise level at the output of the lower and upper correlators in Figure 2. The variance of the correlation of PN code kernel 2Z-1 with the lower sine reference signal multiplied by the noise (as per the demodulation process in the lower leg in Figure 2) is

$$\sigma_L^2 = \frac{1}{N}\left\langle (2Z-1)^2 \sin^2\left(2\pi f t_j\right)\eta^2 \right\rangle = \frac{1}{N}\left\langle \sin^2\left(2\pi f t_j\right)\eta^2 \right\rangle \approx \frac{1}{N}\frac{1}{2}\sigma^2, \quad \text{(A1)}$$

so the standard deviation is

$$\sigma_L \approx \frac{1}{\sqrt{2N}}\sigma . \quad \text{(A2)}$$

For the noise at the output of the upper correlator we have

$$\sigma_U^2 = \frac{1}{N}\left\langle (2Z-1)^2 \cos^2\left(2\pi f t_j\right)\eta^2 \right\rangle = \frac{1}{N}\left\langle \cos^2\left(2\pi f t_j\right)\eta^2 \right\rangle \approx \frac{1}{N}\frac{1}{2}\sigma^2 . \quad \text{(A3)}$$

Here the standard deviation is

$$\sigma_U \approx \frac{1}{\sqrt{2N}}\sigma . \quad \text{(A4)}$$

Z kernel case:

As before we look at the noise level at the output of the lower and upper correlators in Figure 2. The variance of the noise at the variance of the correlation of PN code kernel Z with the lower sine reference signal multiplied by the noise (as per the demodulation process in Fig. 2) is

$$\sigma_L^2 = \frac{1}{N}\left\langle Z^2 \sin^2\left(2\pi f t_j\right)\eta^2 \right\rangle \approx \frac{1}{N}\frac{1}{4}\sigma_L^2 . \tag{A5}$$

so that

$$\sigma_L \approx \frac{1}{2\sqrt{N}}\sigma . \tag{A6}$$

For the upper leg we have

$$\sigma_U^2 = \frac{1}{N}\left\langle Z^2 \cos^2\left(2\pi f t_j\right)\eta^2 \right\rangle \approx \frac{1}{N}\frac{1}{4}\sigma^2 , \tag{A7}$$

so that

$$\sigma_U \approx \frac{1}{2\sqrt{N}}\sigma . \tag{A8}$$

Generally speaking, the noise at the output of the lower and upper correlators ($\eta_L$ and $\eta_U$) will have the same variance, but will generally be statistically independent because they are both the result of two independent correlations.

<u>SNR at the output of the quadrature correlator</u>

For each modulation, we look at the effect of signal and noise through the quadrature correlator. We specifically compare the signal and noise at the correlation peak, since this is how we measure absorption. The output of the quadrature correlator is

$$S_{tot} = \sqrt{\left(S_L + \eta_L\right)^2 + \left(S_U + \eta_U\right)^2} \tag{A9}$$

It's clear that the noise level will be reduced by a significant amount depending on the length of the correlation. We assume the correlation is long enough that the correlation peak of the signal is much bigger than the noise even if we start with a weak signal. Let

$$S = \sqrt{S_L^2 + S_U^2} . \tag{A10}$$

Then

$$
\begin{aligned}
S_{tot} &= S\sqrt{1+2\frac{S_L}{S^2}\eta_L{}'+2\frac{S_U}{S^2}\eta_U{}'+\left(\frac{\eta_L}{S}\right)^2+\left(\frac{\eta_U}{S}\right)^2} \\
&\approx S\left(1+\frac{S_L}{S^2}\eta_L+\frac{S_U}{S^2}\eta_U\right) = S+\sin(\phi)\eta_L+\cos(\phi)\eta_U
\end{aligned}
\tag{A11}
$$

The output noise is then

$$
\eta_{out} \approx \sin(\phi)\eta_L+\cos(\phi)\eta_U \tag{A12}
$$

The variance of the output noise is then

$$
\begin{aligned}
\sigma_{out}^2 &= \left\langle \eta_{out}^2 \right\rangle \approx \left\langle \left(\sin(\phi)\eta_L+\cos(\phi)\eta_U\right)^2 \right\rangle \\
&= \sin^2(\phi)\left\langle {\eta_L}^2 \right\rangle + 2\sin(\phi)\cos(\phi)\left\langle \eta_L\eta_U \right\rangle + \cos^2(\phi)\left\langle {\eta_U}^2 \right\rangle .
\end{aligned}
\tag{A13}
$$

Since $\eta_L$ and $\eta_U$ are statistically independent, $\left\langle \eta_L\eta_U \right\rangle \approx 0$. From our previous results, $\left\langle {\eta_L}^2 \right\rangle \equiv \left\langle {\eta_U}^2 \right\rangle$. Therefore

$$
\sigma_{out}^2 = \left(\sin^2(\phi)+\cos^2(\phi)\right)\left\langle {\eta_L}^2 \right\rangle = \sigma_L^2 \equiv \sigma_U^2 \tag{A14}
$$

The signal to noise is

$$
SNR = \frac{S}{\sigma_{out}} = \frac{S}{\sigma_L} \equiv \frac{S}{\sigma_U} \tag{A15}
$$

Amplitude modulation case with 2Z-1 kernel:

In this case we use 2Z-1 as the kernel in the quatrature correlator and assume a signal with amplitude "a". The signal peak at the lower and upper correlators will be

$$
\begin{aligned}
S_L &= \frac{1}{4}a\sin(\phi), \\
S_U &= \frac{1}{4}a\cos(\phi),
\end{aligned}
\tag{A16}
$$

where is the phase angle between the incoming carrier and the sine reference signal. Using the results of Equations A16 and A2 we have

$$
SNR \approx \frac{S}{\sigma_L} = \frac{\sqrt{S_L^2+S_U^2}}{\sigma_L} = \frac{a\sqrt{N}}{2\sqrt{2}\sigma} \tag{A17}
$$

PSK modulation case with Z kernel:

In this case we use Z as the kernel in the quatrature correlator and assume a signal with amplitude “a”. The signal peak at the lower and upper correlators will be

$$S_L = \frac{1}{4} a \sin(\phi),$$
$$S_U = \frac{1}{4} a \cos(\phi), \tag{A18}$$

where is the phase angle between the incoming carrier and the sine reference signal. Using the results of Equations A18 and A6 we have

$$SNR \approx \frac{S}{\sigma_L} = \frac{\sqrt{S_L^2 + S_U^2}}{\sigma_L} = \frac{a\sqrt{N}}{2\sigma} \tag{A19}$$

This is better than the amplitude modulated case by a factor of $\sqrt{2}$ .

PSK modulation case with 2Z-1 kernel:

In this case we use Z as the kernel in the quatrature correlator and assume a signal with amplitude “a”. The signal peak at the lower and upper correlators will be

$$S_L = \frac{1}{2} a \sin(\phi),$$
$$S_U = \frac{1}{2} a \cos(\phi), \tag{A20}$$

where is the phase angle between the incoming carrier and the sine reference signal. Using the results of Equations A20 and A2 we have

$$SNR \approx \frac{S}{\sigma_L} = \frac{\sqrt{S_L^2 + S_U^2}}{\sigma_L} = \frac{a\sqrt{N}}{\sqrt{2}\sigma} \tag{A21}$$

This is better than the amplitude modulated case by a factor of 2.

<u>SNR for the pure PN code and Hybrid pulse case</u>

In order to make it comparable sine wave signal case for the same average optical power, we take the output signal of 2aZ. For the case of the hybrid pulse we use $4a\sin^2\left(\pi\left(n-1/2\right)P/M\right)Z\left(n\right)$ from the results of Equation 9. The extra factor of 2 comes from equalizing the average optical power for each (assuming the laser amplifier has the headroom to make this possible). Using 2Z-1 as the kernel the correlation peak for either modulation will be

$$S = \frac{N+1}{N} a \approx a \,. \tag{A22}$$

The noise in this case will not have 0 mean because transmitting a pure PN code usually requires a DC coupled receiver or photon counting system. Because the detector will output one sided shot noise, there will always be a bias offset in the noise. The variance of the correlated noise is

$$\sigma_{out}^2 = \frac{1}{N}\left\langle (2Z-1)^2 \left(\eta - \bar{\eta}\right)^2 \right\rangle = \frac{1}{N}\left\langle \left(\eta - \bar{\eta}\right)^2 \right\rangle = \frac{1}{N}\sigma^2 \,. \tag{A23}$$

The signal to noise is then

$$SNR = \frac{S}{\sigma_{out}} \approx \frac{a\sqrt{N}}{\sigma} \,. \tag{A24}$$

This is better than the amplitude modulation case by a factor of $2\sqrt{2}$ . However, unlike the other cases there is a bias offset in the noise. The bias of the correlated noise is

$$\bar{\eta}_{out} = \left\langle (2Z-1)\bar{\eta} \right\rangle = \frac{1}{N_c}\bar{\eta} \,, \tag{A25}$$

where $N_c$ is the bit length (uninterpolated code length) of the PN code. On the other hand, this is small compared to the standard deviation. There is also a small bias term for the other carrier based modulation cases even if the original noise isn't biased. The small terms we threw away in Equation A11 represent a small bias since those noise terms are squared. This is a consequence of the quadrature demodulation, which makes the net signal plus noise always positive. One possible solution to that would be to demodulate the signal with a phase corrected reference carrier and use only one leg of the demodulator. The downside is this may not be simple to do in flight conditions where the range and phase is constantly changing.

## Appendix B: Periodic Gaussian as a Jacobi theta function

The sum represented by Equation 9 may be performed analytically using the Poisson summation formula [27], which states

$$\sum_{n=-\infty}^{\infty} f(n) = \sum_{k=-\infty}^{\infty} \hat{f}(k), \tag{B1}$$

where

$$\hat{f}(k) = \int_{-\infty}^{\infty} f(n) \exp(-2\pi ikn)\, dn, \tag{B2}$$

is the Fourier Transform of *f*. In the case of Equation 1 this becomes

$$\begin{aligned} \hat{f}(k) &= \int_{-\infty}^{\infty} \exp\left((x-nL)^2/2\sigma^2\right) \exp(-2\pi ikn)\, dn \\ &= \sqrt{2\pi}\left|\frac{\sigma}{L}\right| \exp\left(-2\pi^2\sigma^2k^2/L^2\right) \exp(-2\pi kx/L) \end{aligned}, \tag{B3}$$

so that

$$\begin{aligned} F(x) &= \sqrt{2\pi}\left|\frac{\sigma}{L}\right| \sum_{k=-\infty}^{\infty} \exp\left(-2\pi^2\sigma^2k^2/L^2\right) \exp(-2\pi kx/L) \\ &= \sqrt{2\pi}\left|\frac{\sigma}{L}\right| \left[1 + 2\sum_{k=0}^{\infty} q^{n^2} \cos(2\pi kx/L)\right] = \sqrt{2\pi}\left|\frac{\sigma}{L}\right| \vartheta_3(\pi x/L, q) \end{aligned} \tag{B4}$$

**Figure captions**

Figure 1. Baseline instrument block diagram with either band pass or low pass filtered receiver subsystem. Current baseline uses a band pass filter.

Figure 2. The quadrature matched filter correlator used for demodulation is similar to a standard quadrature demodulator except the low pass filter section has been replaced by a matched filter correlation with the reference PN code kernel.

Figure 3. Comparison between cosine (a) and sine wave modulation (b) for amplitude modulation case as in Equation 6. Each modulation waveform produces a different autocorrelation function.

Figure 4. Comparison between autocorrelation function for cosine wave (a) and sine wave case (b) with ML-sequence modulation with 1 cycle per code bit and 8 samples per code bit using a 35 sample delay. Half height width is identical to the pure ML-sequence case.

Figure 5. Comparison between cosine (a) and sine wave modulation (b) for PSK modulation case as in Equation 8. Each modulation waveform produces a different autocorrelation function.

Figure 6. EDFA optical amplifier step response (signal modulation).

Figure 7. Plot of hybrid sine wave pulse laser modulation signal, , using modulation waveform shown in Equation 9, assuming m=1. Note that the received signal in a DC coupled system will be proportional to the modulation signal.

Figure 8. Autocorrelation function for hybrid pulse modulation using a 35 sample delay.

Figure 9. Comparison of frequency content of the filtered vs. unfiltered PN code. Filtered PN code bandwidth is well under the Nyquist sampling rate demonstrating aliasing is no longer an issue.

Figure 10. Comparison between filtered PN code and unfiltered PN code.

Figure 11. Autocorrelation function for filtered vs. unfiltered PN code (filtered case was correlated with unfiltered reference using a 35 sample delay). Off pulse values are zero to within numerical precision.

Figure 12. Filtered PN code and resulting amplitude modulated sine wave.

Figure 13. Frequency spectra of the filtered modulation vs. unfiltered modulation. Modulation bandwidth for the filtered case is under the Nyquist sampling rate demonstrating aliasing is not an issue.

Figure 14. Comparison of autocorrelation functions between filtered modulation case and unfiltered modulation case using a 35 sample delay. Off pulse values are zero to within numerical precision.

Figure 15. Filtered PN code and resulting PSK modulated sine wave.

Figure 16. Frequency spectra of the filtered modulation vs. unfiltered modulation. Modulation bandwidth for the filtered case is under the Nyquist sampling rate demonstrating aliasing is not an issue.

Figure 17. Hybrid pulse modulation with filtered PN code modulation.

Figure 18. Comparison of the frequency spectra between the filtered and unfiltered hybrid pulse case. In this case the filtered PN code modulation bandwidth is under the Nyquest rate, but the unfiltered modulation has a limited bandwidth too.

Figure 19. Comparison between the filtered vs. unfiltered autocorrelation functions for the hybrid pulse case, using a 35 sample delay. Off pulse values are zero to within numerical precision.

**Tables**

Table 1. Comparison of several different modulations comparing Total bandwidth (Total BW), Power band center frequency (PBCF), Signal to noise (SNR) using AM modulation as a reference, and signal background offset (SB Offset). Modulations include pure PN code (PN), Filtered PN code (FPN), Amplitude modulation (AM), PSK with Z kernel (PSK1), Filtered PSK with Z kernel (PSK1), PSK with 2Z-1 kernel (PSK2), and Filterd PSK with 2Z-1 kernel (PSK2).

## Figures

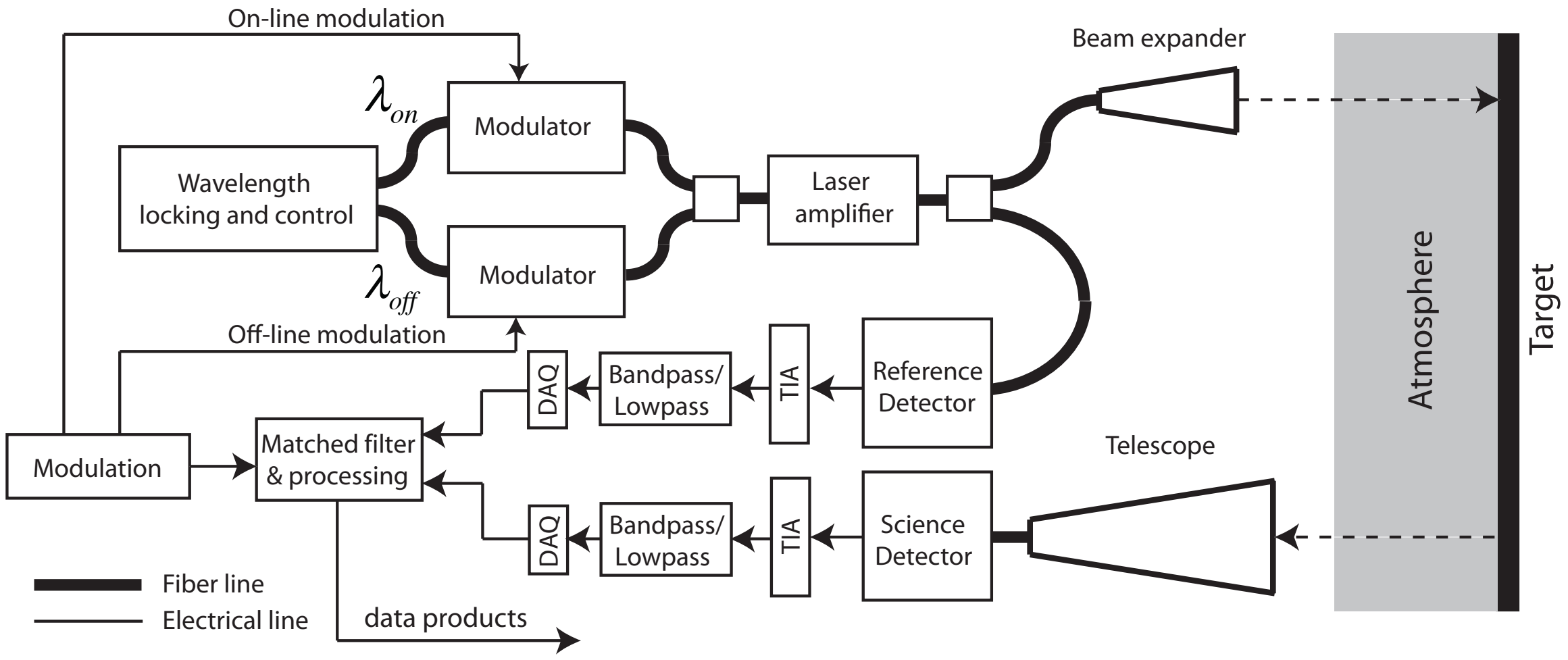

Figure 1. Baseline instrument block diagram with either band pass or low pass filtered receiver subsystem. Current baseline uses a band pass filter.

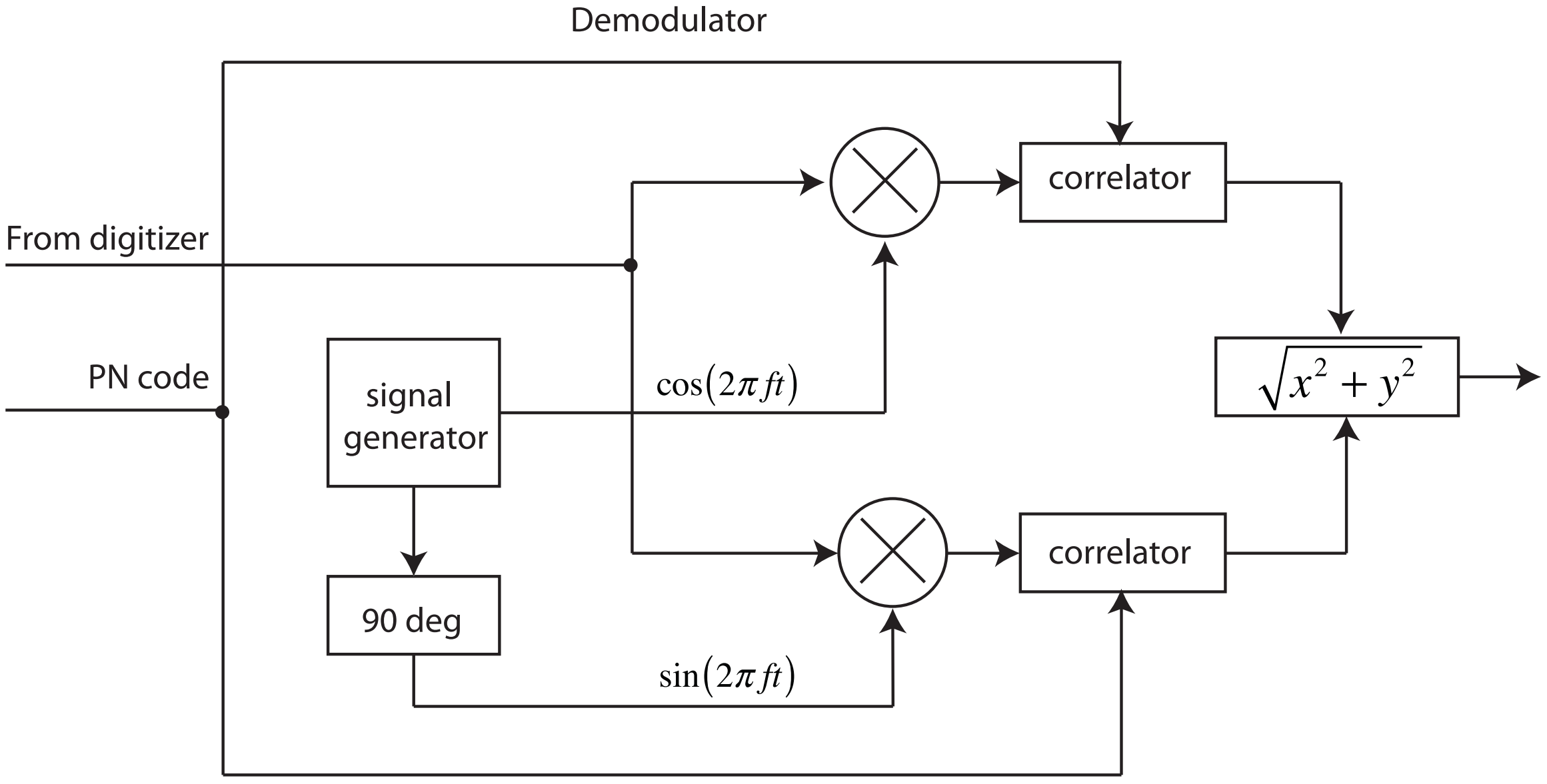

Figure 2. The quadrature matched filter correlator used for demodulation is similar to a standard quadrature demodulator except the low pass filter section has been replaced by a matched filter correlation with the reference PN code kernel.

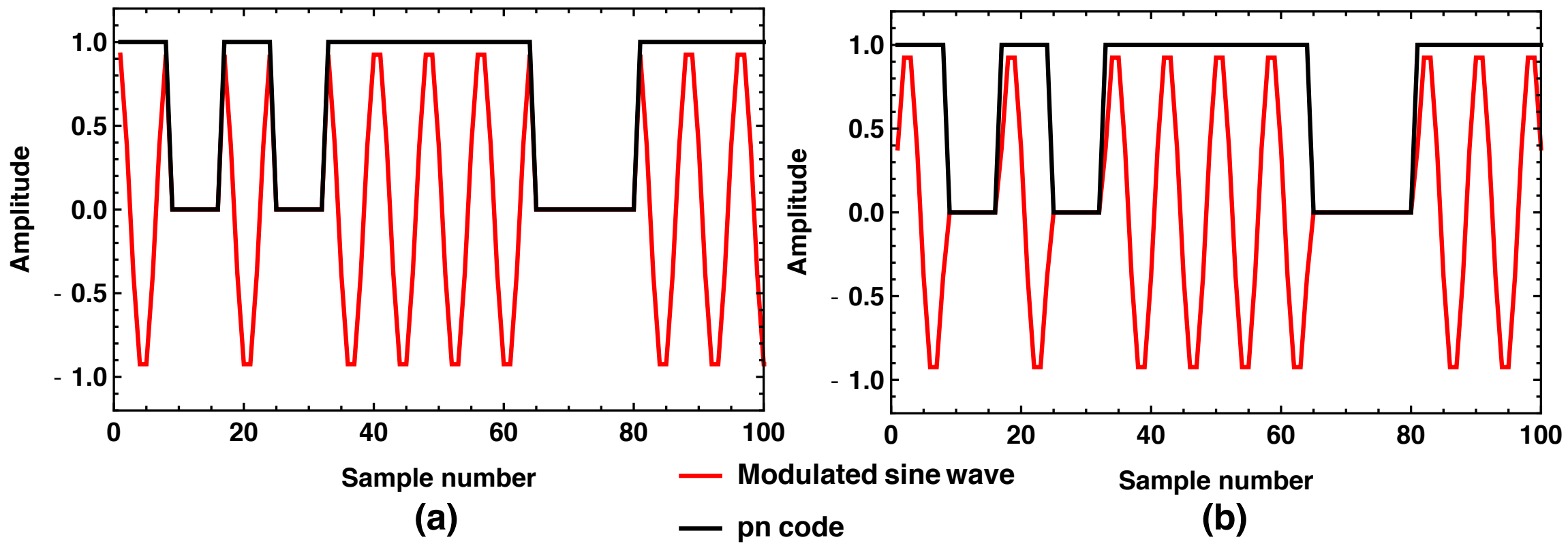

Figure 3. Comparison between cosine (a) and sine wave modulation (b) for amplitude modulation case as in Equation 6. Each modulation waveform produces a different autocorrelation function.

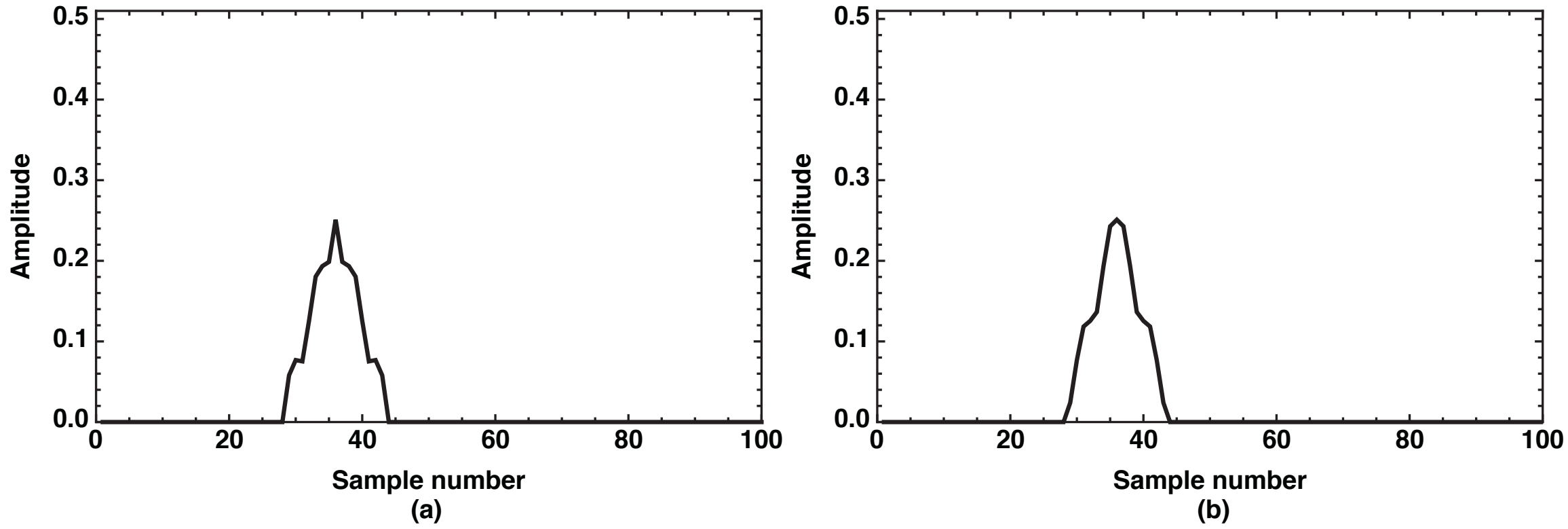

Figure 4. Comparison between autocorrelation function for cosine wave (a) and sine wave case (b) with ML-sequence modulation with 1 cycle per code bit and 8 samples per code bit using a 35 sample delay. Half height width is identical to the pure ML-sequence case.

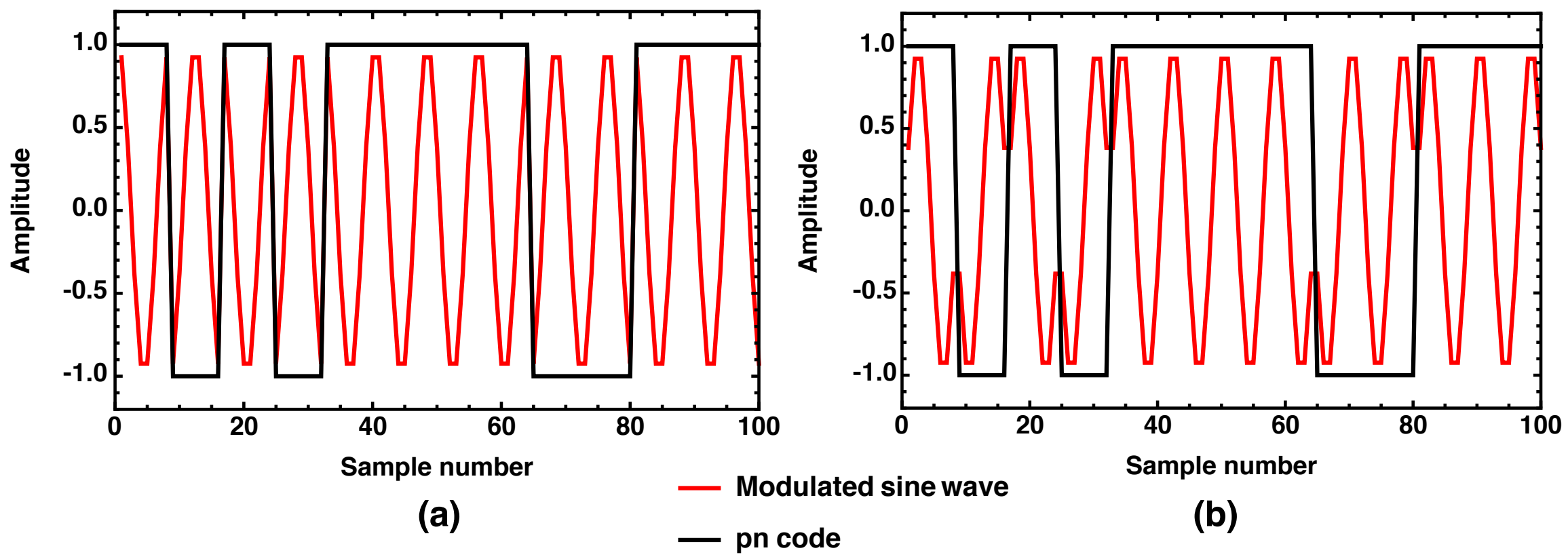

Figure 5. Comparison between cosine (a) and sine wave modulation (b) for PSK modulation case as in Equation 8. Each modulation waveform produces a different autocorrelation function.

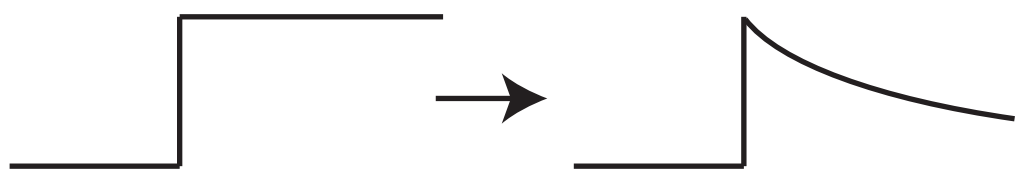

Figure 6. EDFA optical amplifier step response (signal modulation).

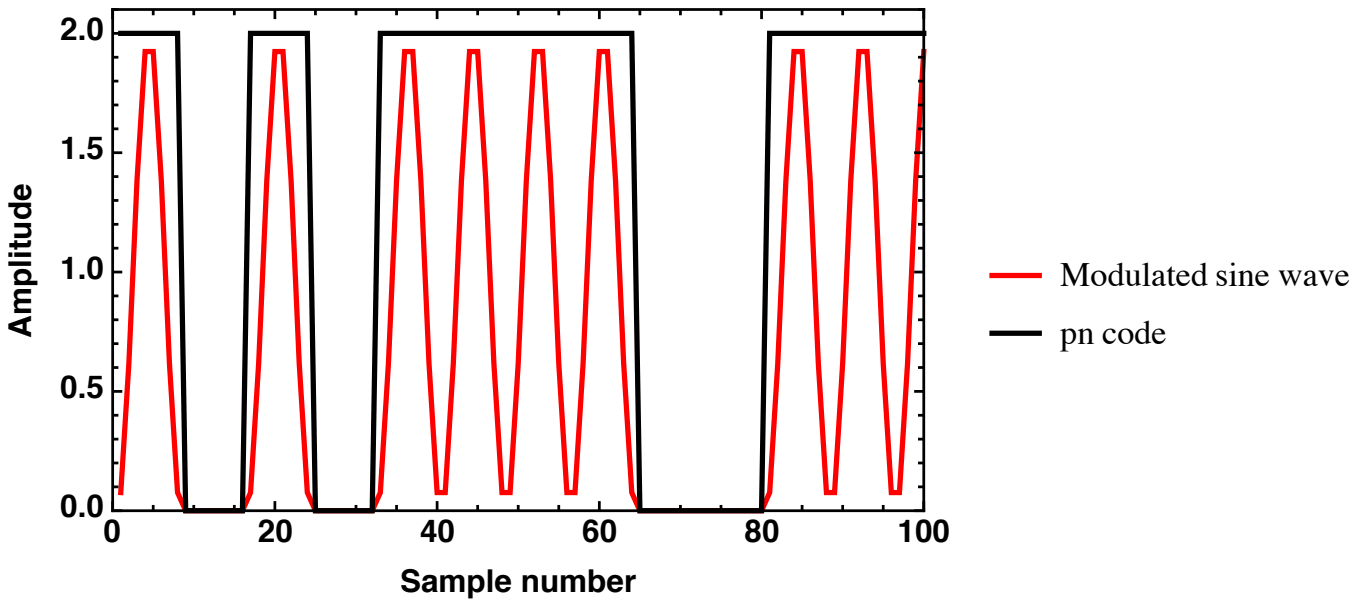

Figure 7. Plot of hybrid sine wave pulse laser modulation signal, $\Lambda$ , using modulation waveform shown in Equation 9, assuming m=1. Note that the received signal in a DC coupled system will be proportional to the modulation signal.

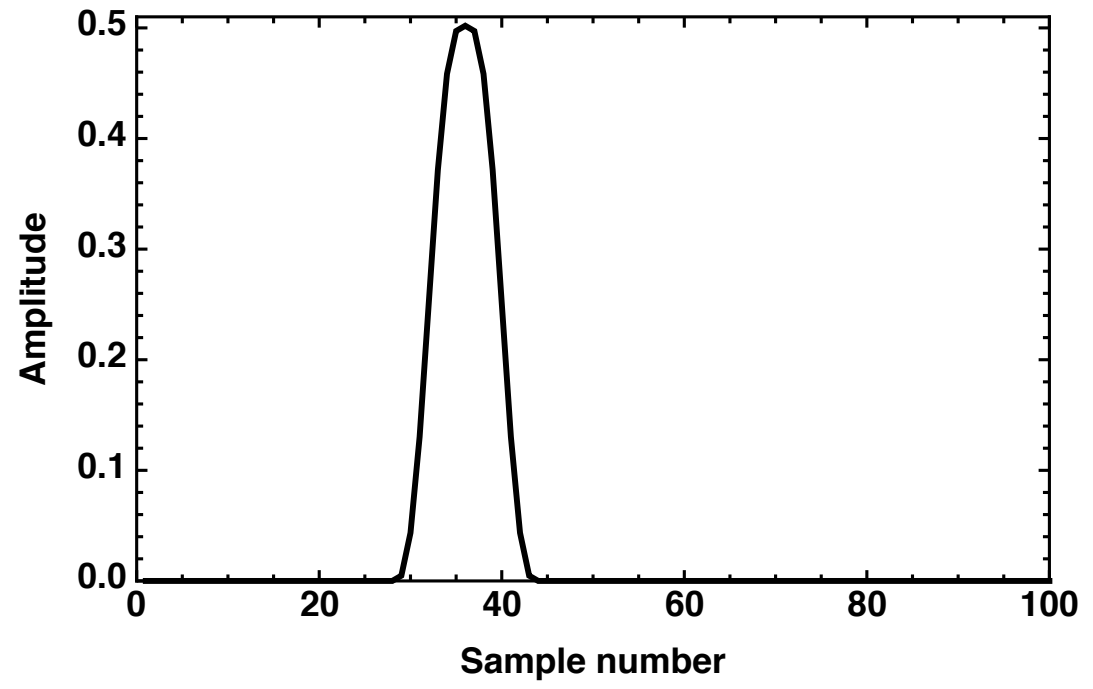

Figure 8. Autocorrelation function for hybrid pulse modulation using a 35 sample delay.

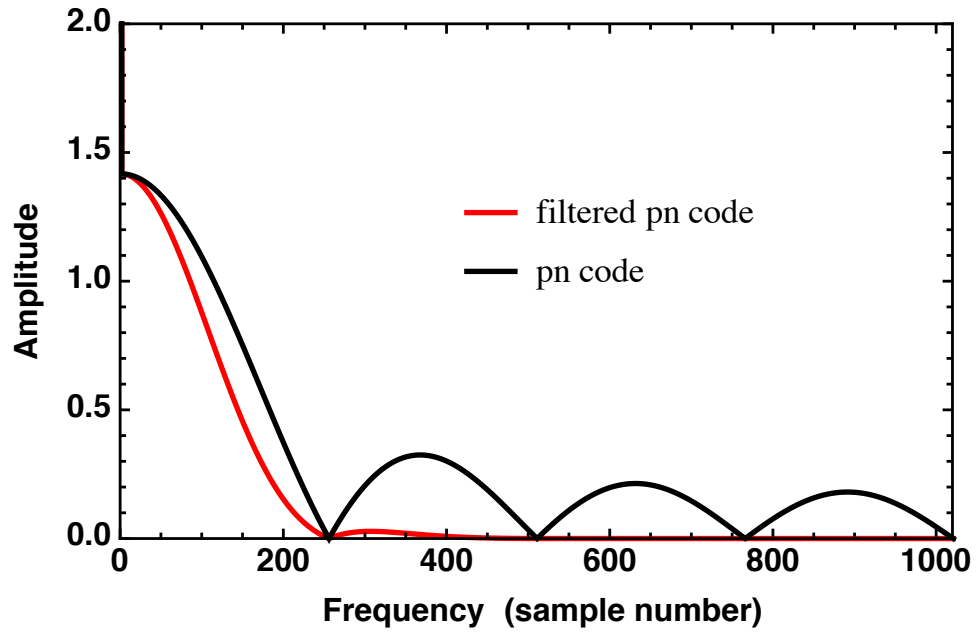

Figure 9. Comparison of frequency content of the filtered vs. unfiltered PN code. Filtered PN code bandwidth is well under the Nyquist sampling rate demonstrating aliasing is no longer an issue.

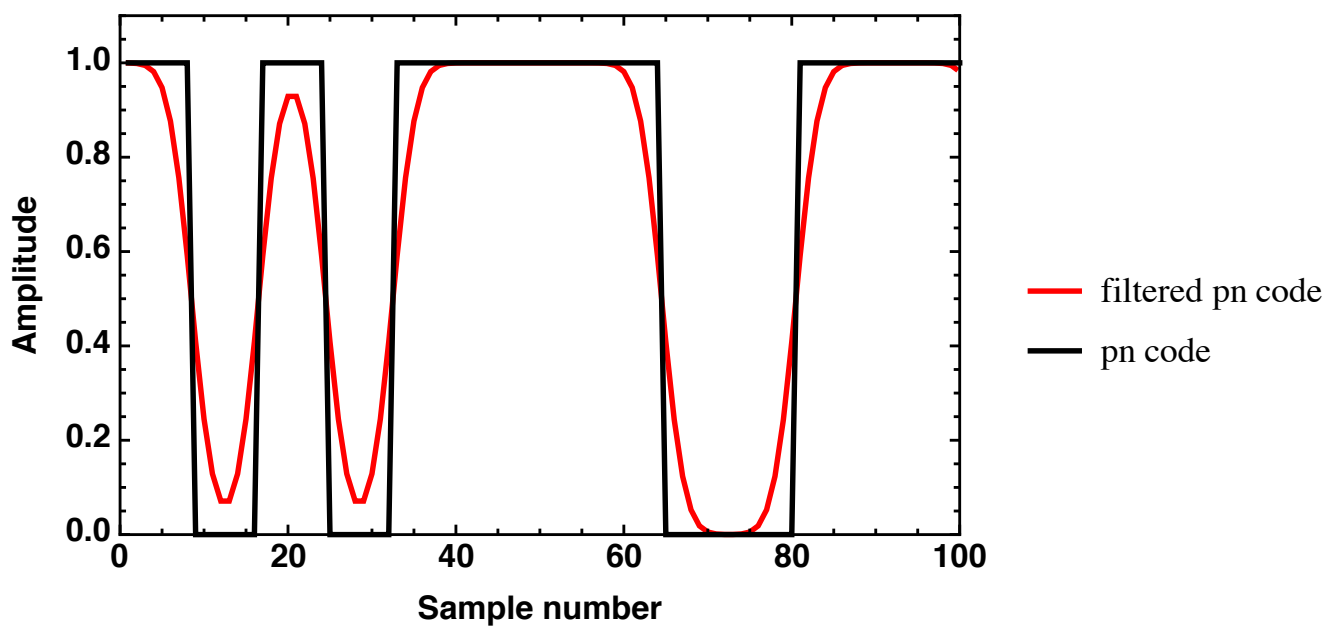

Figure 10. Comparison between filtered PN code and unfiltered PN code.

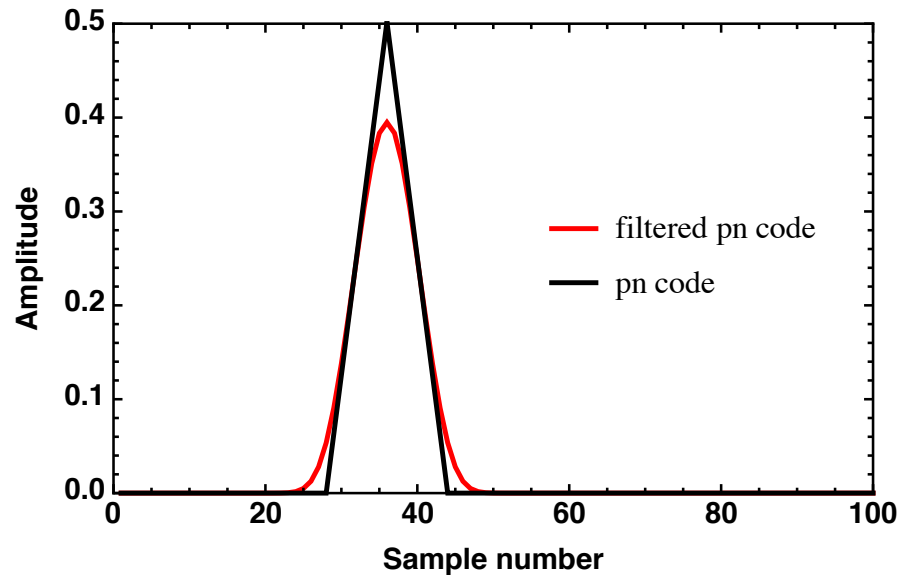

Figure 11. Autocorrelation function for filtered vs. unfiltered PN code (filtered case was correlated with unfiltered reference using a 35 sample delay). Off pulse values are zero to within numerical precision.

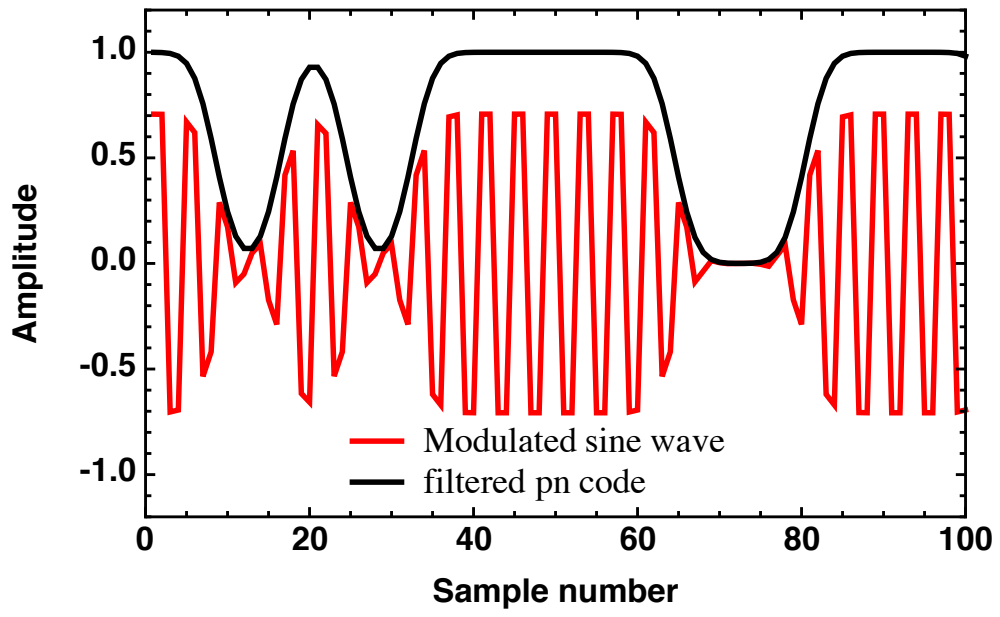

Figure 12. Filtered PN code and resulting amplitude modulated sine wave.

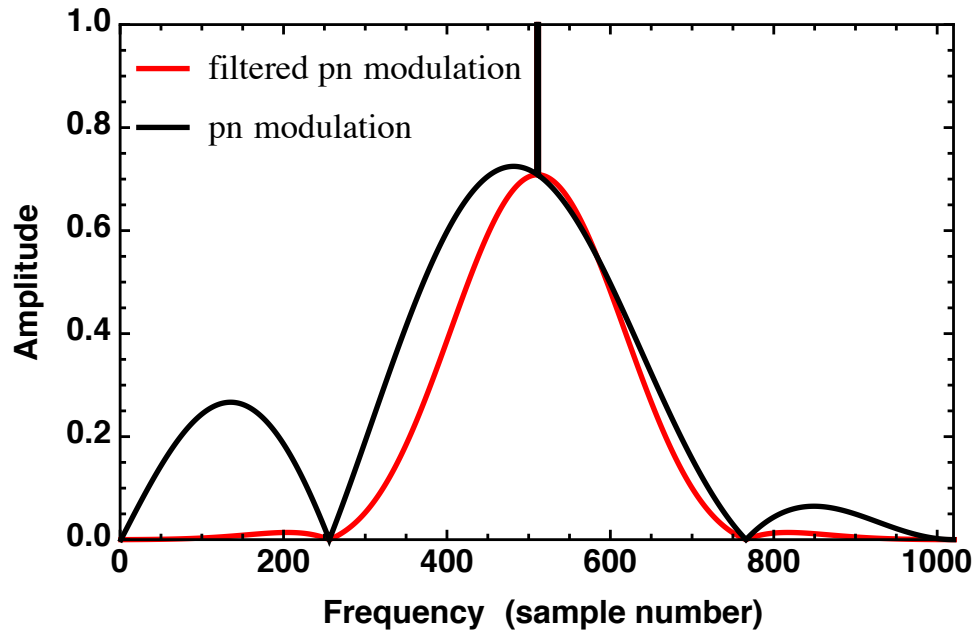

Figure 13. Frequency spectra of the filtered modulation vs. unfiltered modulation. Modulation bandwidth for the filtered case is under the Nyquist sampling rate demonstrating aliasing is not an issue.

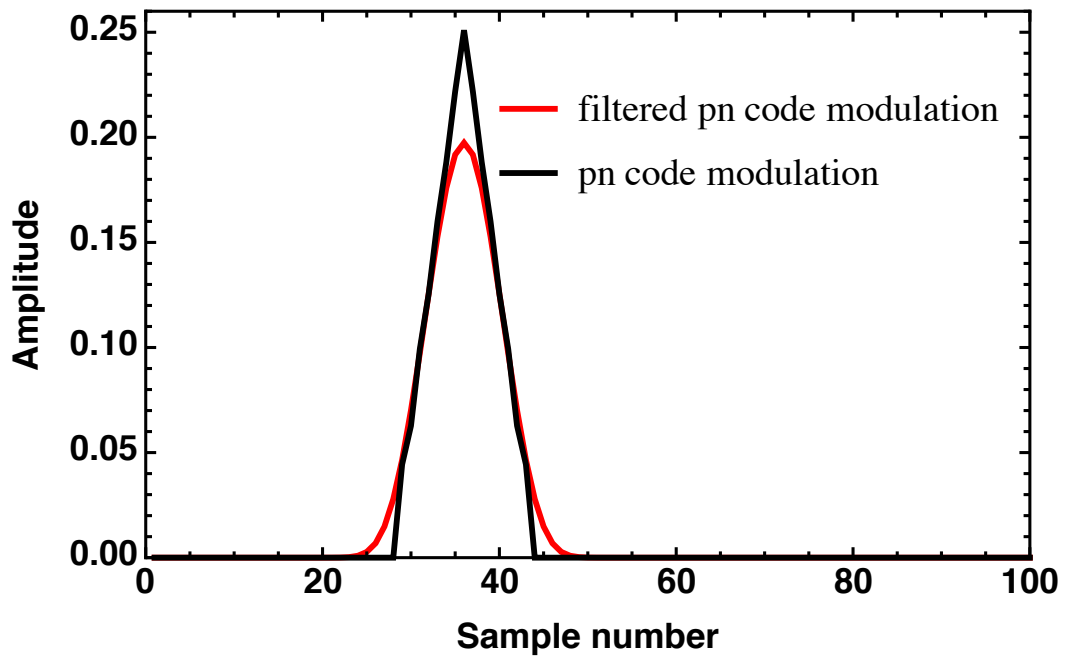

Figure 14. Comparison of autocorrelation functions between filtered modulation case and unfiltered modulation case using a 35 sample delay. Off pulse values are zero to within numerical precision.

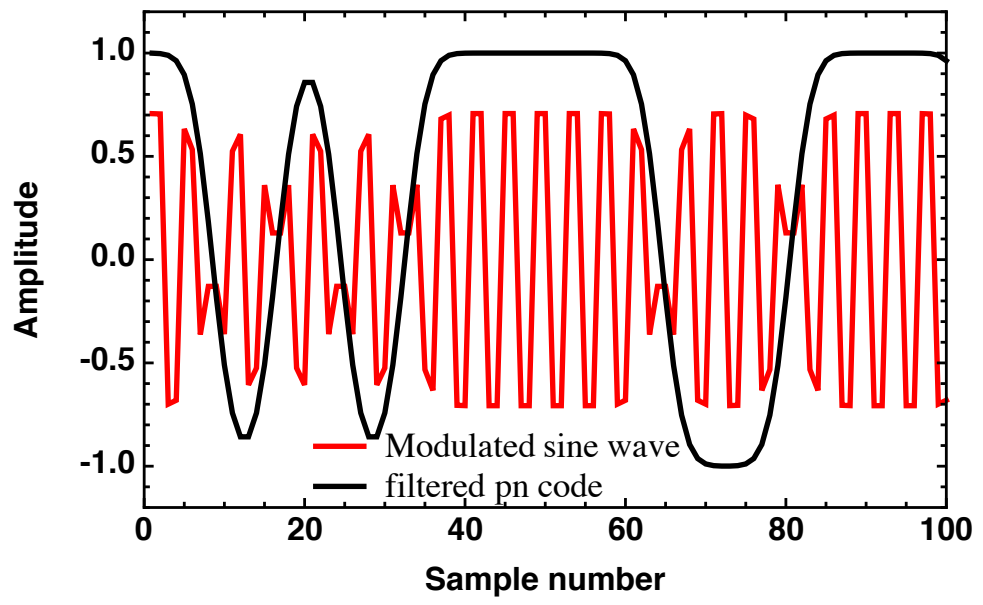

Figure 15. Filtered PN code and resulting PSK modulated sine wave.

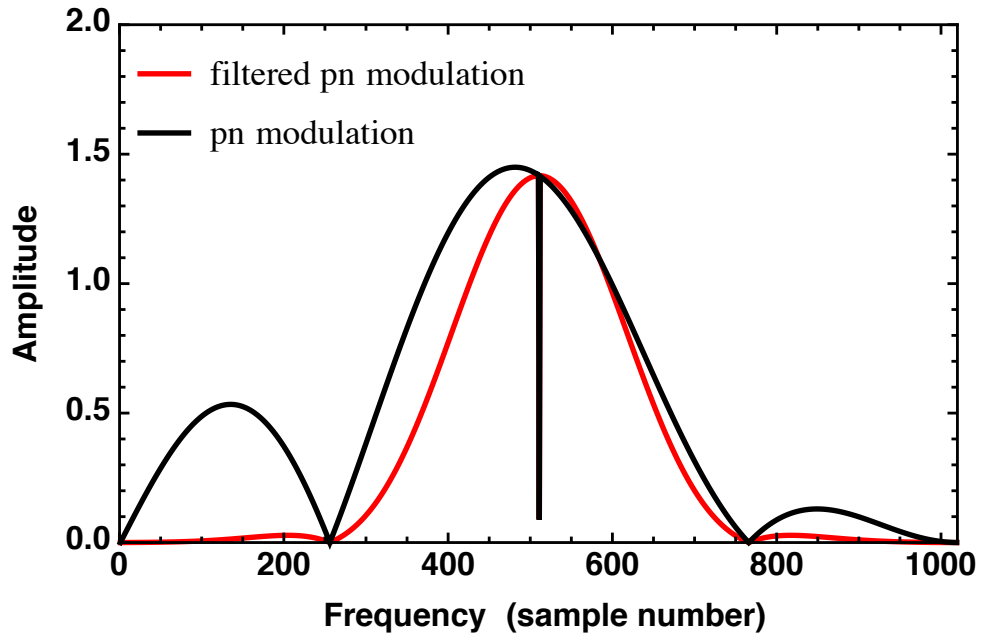

Figure 16. Frequency spectra of the filtered modulation vs. unfiltered modulation. Modulation bandwidth for the filtered case is under the Nyquist sampling rate demonstrating aliasing is not an issue.

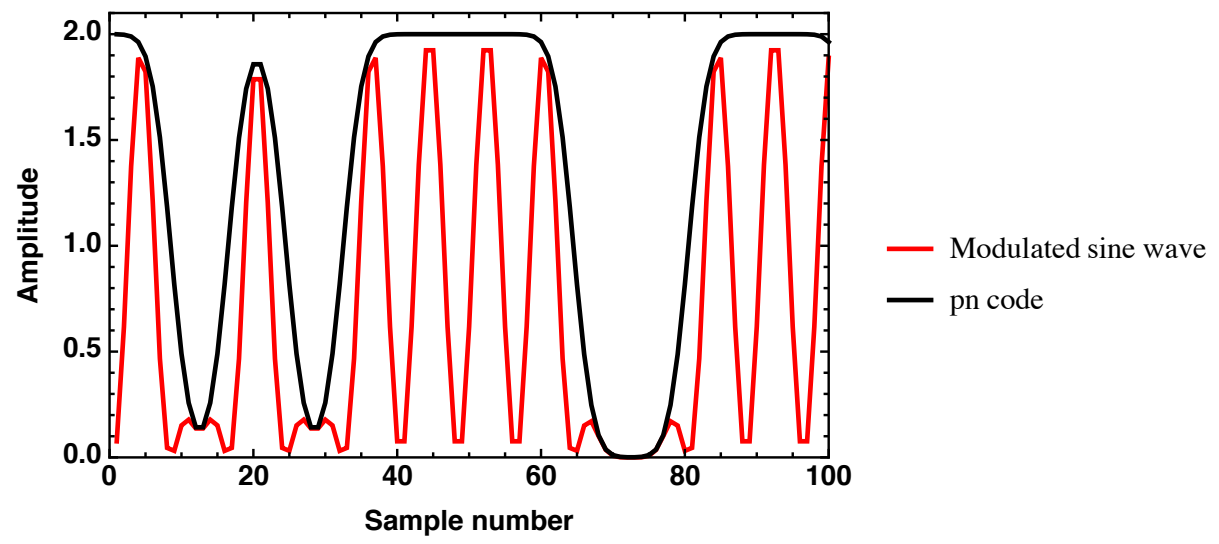

Figure 17. Hybrid pulse modulation with filtered PN code modulation.

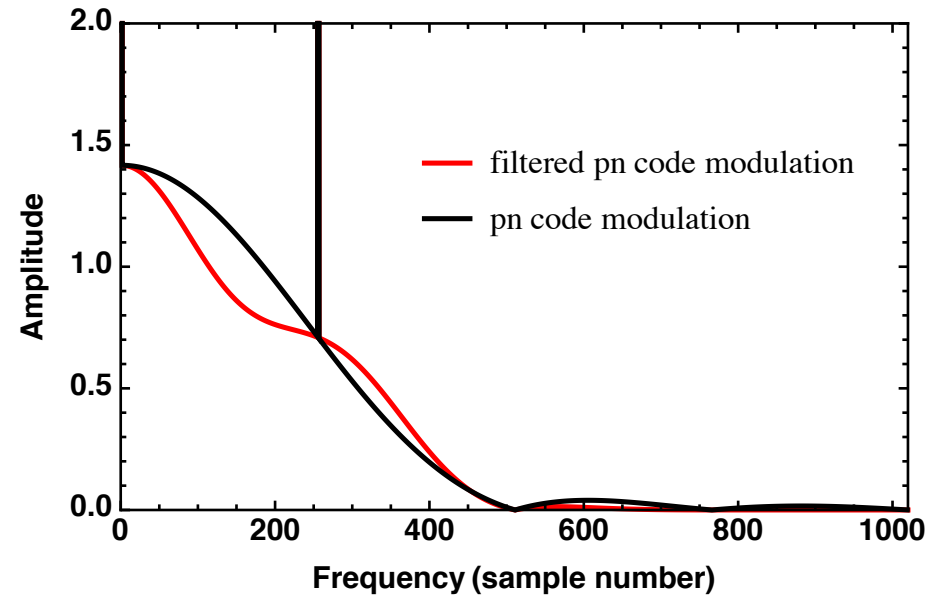

Figure 18. Comparison of the frequency spectra between the filtered and unfiltered hybrid pulse case. In this case the filtered PN code modulation bandwidth is under the Nyquest rate, but the unfiltered modulation has a limited bandwidth too.

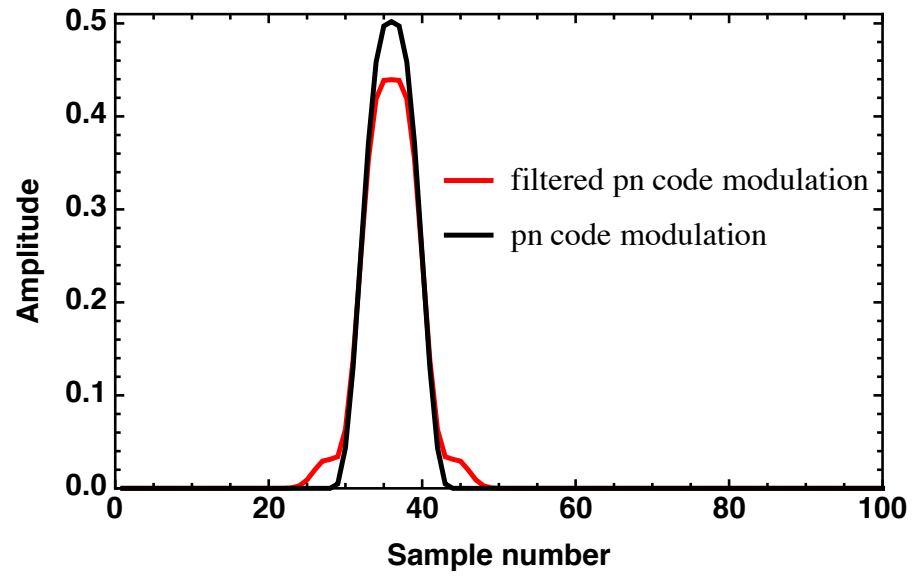

Figure 19. Comparison between the filtered vs. unfiltered autocorrelation functions for the hybrid pulse case, using a 35 sample delay. Off pulse values are zero to within numerical precision.

Table 1. Comparison of several different modulations comparing Total bandwidth (Total BW), Power band center frequency (PBCF), Signal to noise (SNR) using AM modulation as a reference, and signal background offset (SB Offset). Modulations include pure PN code (PN), Filtered PN code (FPN), Amplitude modulation (AM), PSK with Z kernel (PSK1), Filtered PSK with Z kernel (PSK1), PSK with 2Z-1 kernel (PSK2), and Filterd PSK with 2Z-1 kernel (PSK2).

| | Total BW | PBCF | SNR | SB Offset |
|---|---|---|---|---|
| PN | Wide | DC | $2\sqrt{2}$ | No |
| FPN | Variable | DC | $<2\sqrt{2}$ | No |
| AM | Wide | Carrier | 1 | No |
| FAM | Variable | Carrier | $<1$ | No |
| PSK1 | Wide | Carrier | $\sqrt{2}$ | No |
| FPSK1 | Variable | Carrier | $<\sqrt{2}$ | No |
| PSK2 | Wide | Carrier | 2 | Yes |
| FPSK2 | Variable | Carrier | $<2$ | Yes |
| H Pulse | Less wide | DC | $2\sqrt{2}$ | No |
| FH Pulse | Variable | DC | $<2\sqrt{2}$ | No |